# High-definition 3D suspended Archimedean spiral with broadband, spatially extended, and single-handed optical chirality enhancement in Vis−NIR range

*Min Jiang,*[1, 2, 3, *] *Abhik Chakraborty,*[1, 2, 4] *Xiaofei Wu,*[1] *Hark Hoe Tan,*[3] *Jer-Shing Huang*[1, 2, 4, 5, 6, *]

[1]Leibniz Institute of Photonic Technology, Albert-Einstein-Str. 9, 07745 Jena, Germany

[2]Abbe Center of Photonics, Friedrich Schiller University Jena, Albert-Einstein-Str. 6, 07745 Jena, Germany

[3]ARC Centre of Excellence for Transformative Meta-Optical Systems, Department of Electronic Materials Engineering, Research School of Physics, The Australian National University, Canberra ACT 2600, Australia

[4]Institute of Physical Chemistry, Friedrich Schiller University Jena, Helmholtzweg 4, 07743 Jena, Germany

[5]Research Center for Applied Sciences, Academia Sinica, 128 Sec. 2, Academia Road, Nankang District, Taipei 11529, Taiwan

[6]Department of Electrophysics, National Yang Ming Chiao Tung University, Hsinchu 30010, Taiwan

**Abstract**

3D plasmonic structures can provide giant optical chirality (C) in the near field, enabling strong interactions with enantiomers for chiral sensing applications. However, these structures face several limitations, including narrow operational bandwidth constrained by resonance, flipping handedness of C around the resonant frequency, spatially confined distribution of C, and difficulty in trapping enantiomers. Our numerical predictions reveal that a 3D plasmonic Archimedean spiral (AS) exhibits spectrally broadband, spatially extended and single-handed near-field C in visible-to-near-infrared range. However, realizing highly deterministic 3D AS remains challenging. We develop an effective fabrication strategy that combines focused ion beam milling and template-stripping method to realize high-definition 3D AS. Furthermore, we experimentally characterize the fabricated structure's far-field chiroptical behavior to confirm the predicted optical response. Owing to its conical hole-like geometry, 3D AS can potentially function as a sieve for trapping enantiomer-functionalized nanoparticles in the hot zone with enhanced C for sensitive broadband chiroptical detection.

**Keywords:** Archimedean spiral, optical chirality enhancement, circular dichroism, chiral sensing, 3D plasmonics

## Introduction

A molecule is described as chiral if its mirror image cannot be superimposed on itself.[1] Enantiomers, a pair of chiral isomers with oppositely handed spatial arrangement of atoms, can exhibit different or even opposite biochemical functions due to their non-superimposability in relation to each other. Thus, distinguishing enantiomers is crucial in pharmaceutical and biomolecular sciences to eliminate the unwanted side effects. Circular dichroism (CD) spectroscopy reveals the absorption difference of chiral molecules when exposed to oppositely handed circularly polarized light (CPL), enabling the identification of their chiral state.[2–4] However, the intrinsic chiroptical response of enantiomers is inherently weak due to the mismatch between the dimension of the molecular chiral domains and the wavelength of light in the visible-to-near-infrared (Vis−NIR) range. Also, the sensitivity of CD spectroscopy is hindered by the difficulty in preserving the degree of circular polarization along the optical path. This makes CD detection of chiral molecules at low concentration very challenging.

Absorption CD of a molecule stems from the interference of the induced electric moments and magnetic moments[5] and is mainly proportional to the optical chirality (C) of the illuminating light (see mathematical deviation in Support Information).[6,7]

$$A^{\pm} = \frac{2}{\varepsilon_0}(\omega U_e \alpha'' \mp C G'') \qquad (1)$$

$A^{\pm}$ are the absorption rate for left-handed CPL (+, L-CPL) and right-handed CPL (−, R-CPL), $\varepsilon_0$ is the permittivity of vacuum, $\omega$ is the angular frequency, $U_e = \frac{\varepsilon_0}{4}|\mathbf{E}|^2$ is the time-average energy density of the electric field, $\alpha''$ is the imaginary part of electric polarizability, $C = -\frac{\varepsilon_0 \omega}{2}\text{Im}(\mathbf{E}^* \cdot \mathbf{B})$, $G''$ is the imaginary part of mixed electric-magnetic dipole polarizability, $\mathbf{E}$ and $\mathbf{B}$ are the electric and magnetic field, respectively. For propagating plane waves, maximum C is found in perfect CPL. However, using plasmonic nanostructures could generate optical near fields with enhanced C exceeding that of CPL.[8–10] This is achievable through: (1) amplifying the electromagnetic fields by orders of magnitude via subwavelength mode confinement at the hot spots,[11–15] (2) optimizing the polarization and relative phase between $\mathbf{E}$ and $\mathbf{B}$ to maximize $\text{Im}(\mathbf{E}^* \cdot \mathbf{B})$.[16–18] The local enhancement in C is defined as optical chirality enhancement ($\hat{C}$),

$$\hat{C}^{\pm} = \frac{C^{\pm}_{\text{local}}}{|C^{\pm}_{\text{CPL}}|} \qquad (2)$$

$C^{\pm}_{\text{local}}$ and $C^{\pm}_{\text{CPL}}$ are the values of C obtained for L-CPL (+) and R-CPL (−) illumination with and without the nanostructure, respectively. The sign of $C_{\text{CPL}}$ is not considered during the calculation of $\hat{C}$ to easily distinguish between the handedness of the local field in the vicinity of the nanostructure under different illumination chirality. Therefore, higher $\hat{C}$ means higher CD, and vice versa.[19–22]

3D plasmonic spirals interact strongly with CPL due to their structural congruence with the electric field vector,[23–26] enabling phase-matched resonant coupling when handedness aligns. This drives exceptional $\hat{C}$ due to the conformity between the helicity of field and structure, and long-range chiral field propagation along the structure's axis. However, the enhancement is based on plasmonic resonance, restricting the operational bandwidth. Additionally, many designs exhibit near-field $\hat{C}$ with alternating handedness in closely adjacent spatial and/or spectral regimes, effectively cancelling the overall enhancement effect.[18,27,28] To overcome these limitations, the ideal design needs to achieve not only single-handed and broadband $\hat{C}$ covering several molecular transition bands, but also needs to enforce it over a large accessible region where analytes can stay and experience the aforementioned $\hat{C}$.

To the best of our knowledge, our previous study[29] remains the only demonstration of broadband $\hat{C}$. This is achieved in a 3D Si$_3$N$_4$/Au Archimedean spiral (AS) across the 2−8 μm wavelength range. Focused ion beam (FIB) milling of the AS pattern into a free-standing Au/Si$_3$N$_4$ bilayer film with intrinsic stress imbalance stretches out the freshly milled planar AS vertically, thereby forming a 3D AS structure. However, the resulting AS looks more like a tilted blunt spiral with uncontrollable and little out-of-plane stretch, and therefore fails to meet the desired conical shape. An alternative fabrication technique, nano-kirigami, creates 3D plasmonic structures by combining FIB milling and ion irradiation to buckle planar interconnected features into 3D geometries such as 3D pinwheel arrays.[30] However, in this method, the deterministic out-of-plane displacement of central region remains limited. Subsequent "kissing-loop" nano-kirigami overcomes this by breaking the central connection of the feature and employing two isolated short arcs, thereby achieving 3D split-ring structure with a sharp tip.[31] While this approach resolves the issue of restricted central displacement, it remains incompatible with structurally complex designs like 3D AS which requires continuously varying geometrical parameters in 3D to achieve true conical helicity.

In this work, we develop a fabrication strategy combining FIB and template stripping to overcome the aforementioned challenges in fabrication and demonstrate suspended 3D AS that exhibits spectrally broadband and spatially extended single-handed $\hat{C}$ across the Vis−NIR regime. With the developed fabrication method, we obtain high-definition and suspended 3D right-handed AS (r-AS) structures with a sharp apex. The fabricated structure is characterized by transmission CD spectroscopy. The corresponding dissymmetry factor (g-factor) shows preservation of sign across a broad working wavelength region, confirming the broadband single-handed chiroptical response.

**Results and discussion**

To improve the conical stability of the structure, we design a two-arm 3D AS instead of a single-arm AS. As shown in **Figure 1A** and **1B**, r-AS consists of two spiral arms with six full turns. Two arms follow contours described by $r(\theta_1) = \Delta r \times \theta_1/2\pi$ and $r(\theta_2) = \Delta r \times \theta_2/2\pi$, where $r$ (nm) is the distance from the common spiral center, $\Delta r = 150$ nm is the radial increment per $2\pi$ turn, and $\theta_1, \theta_2$ are the in-plane azimuthal angles describing the angular trajectory of the arms ($\theta_1 = [0, 12\pi]$, $\theta_2 = [\pi, 13\pi]$). The vertical gap between adjacent spiral turns is 30 nm. To strengthen the structure, the two arms are connected with 30 nm wide bars that are placed every 60º along the spiral. The top surface of the Au film is defined as z = 0 nm. The r-AS reaches a height of 1280 nm at the outer surface of the apex and z = 1200 nm at the inner tip.

The r-AS' base diameter at the bottom surface of the Au film is 2000 nm. An example of the fabricated r-AS with this optimal design is shown in **Figure 1C**.

Considering its conical hole-like geometry, suspended 3D r-AS can potentially work as a sieve to mechanically trap nanoparticles within its inner cavity. These particles can be surface-functionalized with chiral molecules and thus serve as a vehicle to bring chiral targets to the hot zone for achieving chiral sensing at low concentration. Considering the diameter of such kind of particles, the functional region of interest is inside the inner hollow cavity with a span of 400 nm in the z-direction (800−1200 nm), illustrated in the inset of **Figure 1A**.

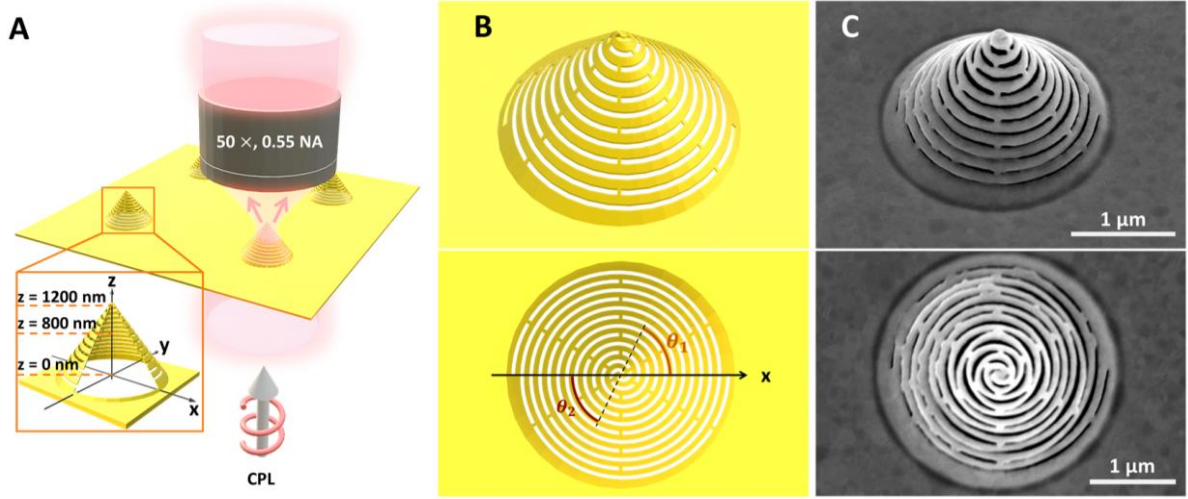

**Figure 1**. (A) Schematic of 3D r-AS array under CPL illumination at surface normal incidence from bottom. The inset shows the coordinate system used in this work. The z = 0 nm plane is defined as the top surface of the Au film (thickness: 80 nm). The inner surface of the tip is at z = 1200 nm. The diameter of the base of the r-AS measured at the bottom surface of the Au film is 2000 nm. (B) Design of the 3D r-AS shown in tilted (upper panel) and top (lower panel) view. The black dashed line marks the in-plane azimuthal angles $\theta_1$ and $\theta_2$ relative to the x-axis, describing the angular trajectories of the two spiral arms ($\theta_1 = [0,12\pi]$, $\theta_2 = [\pi, 13\pi]$). $\theta_1$ counts from the positive x-axis while $\theta_2$ counts from the negative side. (C) scanning electron microscope (SEM) image of an example of the fabricated r-AS under tilted (upper panel) and top (lower panel) views.

The optical response of the 3D r-AS, especially its potential for chiral sensing, is numerically investigated without any nanoparticle. A 2D monitor in the y-z plane at x = 0 nm records the electric field intensity distribution across the functional region. Under R-CPL illumination (**Figure 2A**), it displays a strong intensity enhancement near the tip, persisting across wavelengths of 600−850 nm. However, L-CPL illumination (**Figure 2B**) does not lead to such enhancement, indicating excitation chirality-dependent near-field intensity enhancement.[32] Moreover, across the functional region away from the tip, the intensity enhancement under R-CPL is higher than that under L-CPL. Considering the AS as a plasmonic waveguide with a tapered helical geometry, this effect can be interpreted as the excitation chirality-dependent coupling of light to plasmon polaritons.[33,34] The chirality-dependent coupling of light is dependent on the match or the mismatch between the handedness-dependent sign of the topological charge of the AS and the handedness-dependent sign of the topological charge of the incident CPL.[29] When an AS of a certain handedness is excited by CPL of the favorable handedness, a guided field propagates along the AS towards the tip, resulting in a focused field. In contrast, when the handedness of the structure and that of the incident CPL are mismatched, the field gets distributed to various locations in the AS depending on the wavelength and the local spiral radius. This behavior arises because the AS can be viewed as a set of staggered split rings, where the resonant wavelength depends on ring size.[29] In our case, the combination of r-AS and R-CPL (matched case) leads to efficient light-matter coupling, constructive interference and intensity enhancement near the tip. Conversely, for the mismatched case, the field confinement is poor with no significant intensity enhancement near the tip.

Next, we quantitatively evaluate near-field chiroptical behavior by calculating $\hat{C}$ using the same 2D monitor and eq 2. Notably, under R-CPL illumination, we see single-handed $\hat{C}$ across both the spectral and spatial domains with the highest value reaching 25 (**Figure 2C**). In contrast, L-CPL illumination leads to smaller value of $\hat{C}$ and both handedness distribution of $\hat{C}$ over the entire spectral and spatial domains of operation (**Figure 2D**). Broadband $\hat{C}$ stems from the fact that the 3D r-AS works as a horn antenna that exhibits broadband optical response with self-focusing of the near field to the tip of the spiral.[29,32] The clear difference in intensity enhancement and $\hat{C}$ support our proposal of using 3D r-AS to induce single-handed, broadband and spatially extended $\hat{C}$ within the functional region.

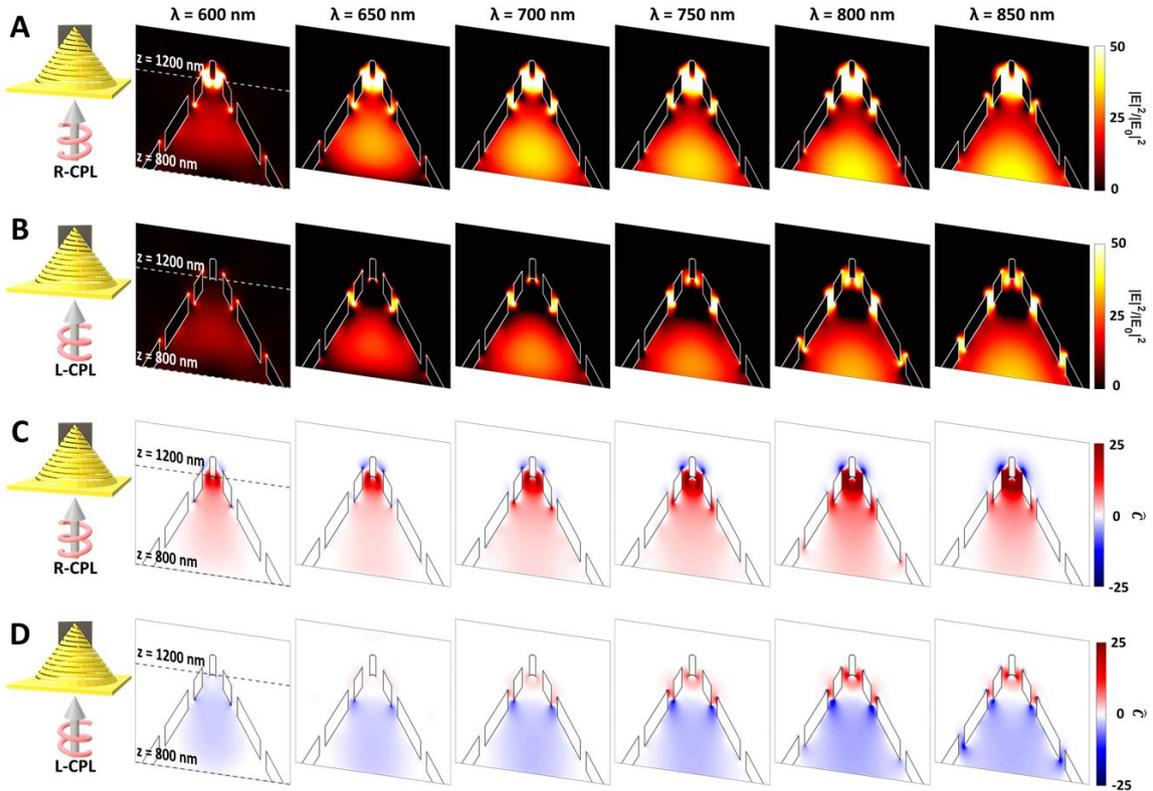

**Figure 2.** Near-field intensity distribution of 3D r-AS in the y-z plane (x=0 nm) at various wavelengths under (A) R-CPL and (B) L-CPL excitation. Cross-section distribution of $\hat{C}$ on y-z plane at various wavelength under (C) R-CPL and (D) L-CPL excitation.

To evaluate $\hat{C}$ of the 3D r-AS across the continuous wavelength range, we use a line monitor (black line in **Figure 3A**) along the z-axis. Under R-CPL excitation (**Figure 3A**), a strong and single-handed Ĉ is observed across $\lambda = 600-850$ nm within the entire functional region, with the highest Ĉ of 25 near the tip. Under L-CPL excitation (**Figure 3B**), both handedness of Ĉ are excited across the spectral and spatial range of interest. Here, the overall Ĉ is significantly reduced. Since linearly polarized light is easier to control in practice, we also explore the possibility of yielding significant Ĉ in the 3D r-AS excited by linearly polarized light (**Figure 3C**). **Figure 3C** suggests that the resulting Ĉ distribution is indeed significant but more confined near the tip compared to **Figure 3A** and **3B**. This implies the robustness and reliability of 3D r-AS under excitation with varying polarization state. For comparison, the corresponding behavior of a left-handed AS under three excitation conditions is numerically investigated and shown in **Figure S1** in Supporting Information.

Considering the case of a functionalized nanoparticle trapped inside the 3D r-AS near the tip in the proposed sieve application, a silica sphere of 100 nm diameter is added in the simulation. As seen in **Figure S2** in Supporting Information, the presence of the silica sphere inside the structure does not cause noticeable difference in Ĉ distribution compared to **Figure 3**, confirming the feasibility of the AS' application as a sieve for chiral sensing at low concentrations.

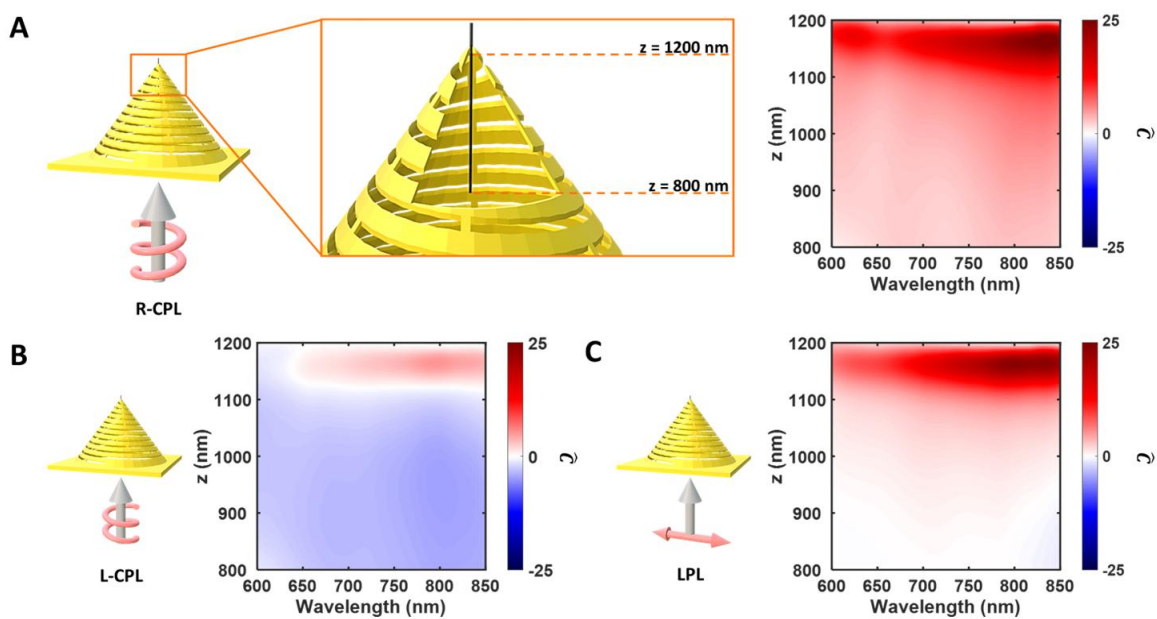

**Figure 3.** Ĉ of the 3D r-AS calculated from the line monitor oriented along the central z-axis (black line) under (A) R-CPL, (B) L-CPL, and (C) linearly polarized light excitation.

A complete pictorial description of the fabrication method is provided in **Figure 4A**. It involves three main steps: (a) creating embedded Au cone array, (b) transferring Au cone array, and (c) patterning r-AS structures. The process begins with the milling of an inverse cone array into a Si substrate using Ga$^+$ FIB. The milled Si substrate is then coated with a 5 nm-thick SiO$_2$ and subsequently by an 80 nm-thick Au film. The Au film with the hollow cone array is transferred to a commercially bought Si substrate with square aperture using polymethyl methacrylate (PMMA). The cleaned Au cone array suspended over the square aperture is subjected to Ga$^+$ FIB to fabricate the two-arm r-AS pattern. More details regarding this can be found in Methods.

**Figure 4B** presents the SEM image of the transferred Au cone array while **Figure 4C** shows the milled 3D r-AS. The resulting structures closely match the design specifications. Importantly, no deformation or collapse is observed post-FIB milling, highlighting the reliability and reproducibility of our approach for the fabrication of complex 3D structures with nanoscale geometrical features. An SEM image of the complete 6×6 array of fabricated structures is shown in **Figure S3** in Supporting Information.

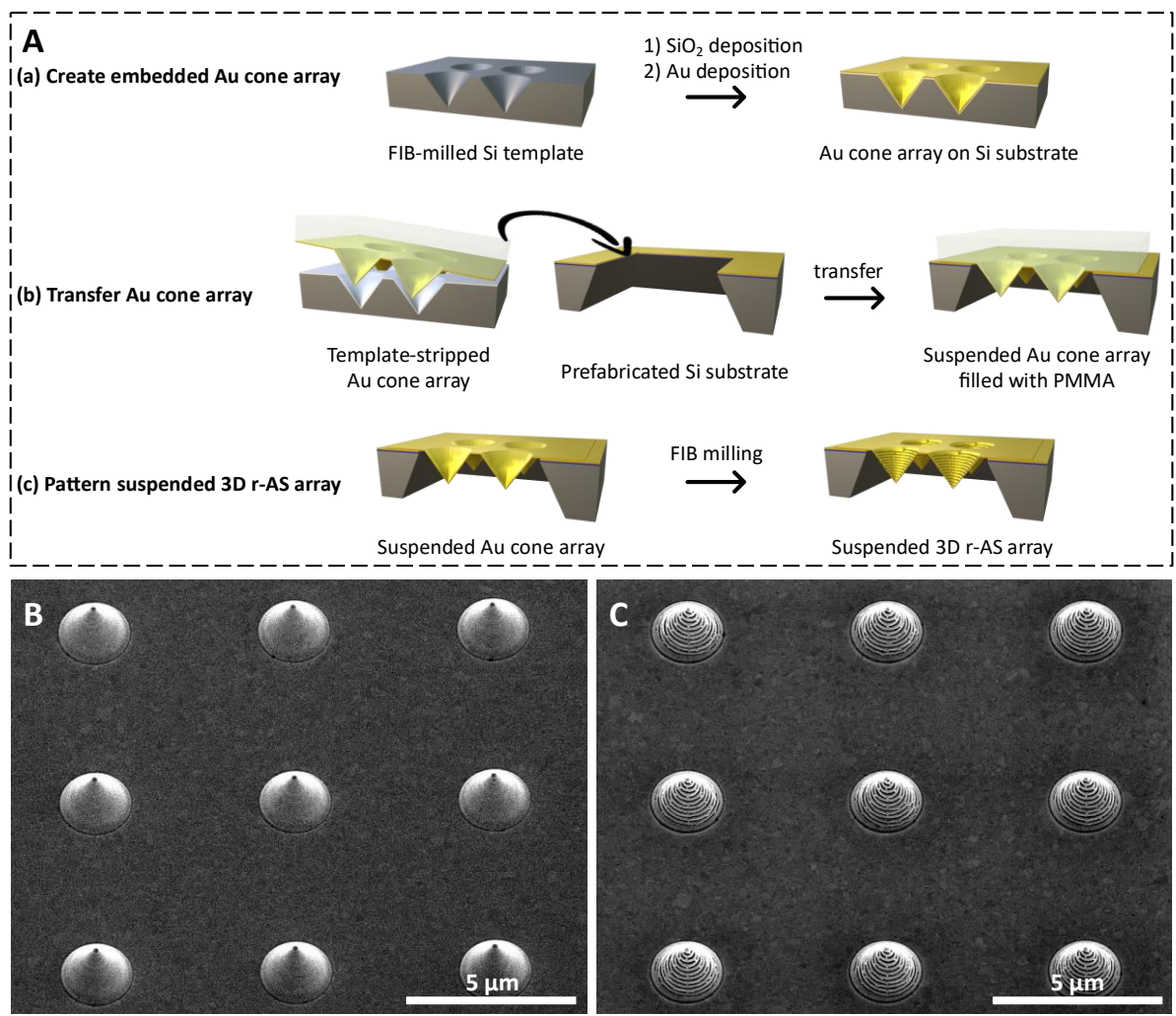

**Figure 4.** (A) The schematic illustration of the fabrication procedure of the 3D suspended r-AS array. (B) SEM image of 3D suspended Au cone array (tips pointing upwards, 30° tilted view) after removing PMMA. (C) SEM image of 3D suspended r-AS array (tips pointing upwards, 30° tilted view).

To confirm the broadband and single-handed chiroptical response of the 3D r-AS, the simulated transmission spectra of a single r-AS and the experimental transmission spectra under L-CPL (red) and R-CPL (black) illumination are investigated, shown in **Figure 5A**. The transmission measurement is done by a home-built microscope setup detailed in Methods and Supporting Information (**Figure S4**). Both simulation and experimental results suggest higher transmission under L-CPL illumination compared to R-CPL. The experimental spectra, however, exhibit lower transmission compared to simulation, which can be attributed to losses arising from structural imperfections such as surface roughness of the r-AS and non-uniform gap between adjacent spirals in the same r-AS. Despite the smooth line shape of simulated spectra, minor features are observed in the experimental spectra. This can be attributed to the fact that, in the experiment, the whole array is excited under collimated illumination. Therefore, even though spatial filtering is done in the conjugated image plane to collect the transmission from one r-AS, there might be contribution from the interference of the transmitted light from periodic structures. In the simulation, one structure is considered. To verify the effect of this difference between experimental and simulation conditions, we also simulate transmission spectra for the case where the entire 6×6 array is illuminated (see **Figure S5** and associated simulation details in Supporting Information). In this case, undulating spectral features are reproduced in the simulated transmission spectra. The efficiency of the chiroptical response is quantified using the g-factor ($g$), defined as:

$$g = \frac{T_{L-CPL} - T_{R-CPL}}{T_{L-CPL} + T_{R-CPL}} \times 2 \qquad (3)$$

Both simulated $g_{sim}$ and experimental $g_{exp}$ are calculated using eq 3. **Figure 5B** displays good agreement between $g_{sim}$ and $g_{exp}$ over the entire spectral range, exhibiting pronounced chiroptical behavior with $g_{exp}$ reaching a maximum of 0.42 at 798 nm. This agreement indicates that the fabricated r-AS closely matches the numerical model, which is remarkable considering the complexity of the structure. Besides, $g_{sim}$ and $g_{exp}$ provide compelling evidence for the broadband chiroptical behavior in r-AS, as indicated by the consistent sign preservation over the entire spectral regime of operation.

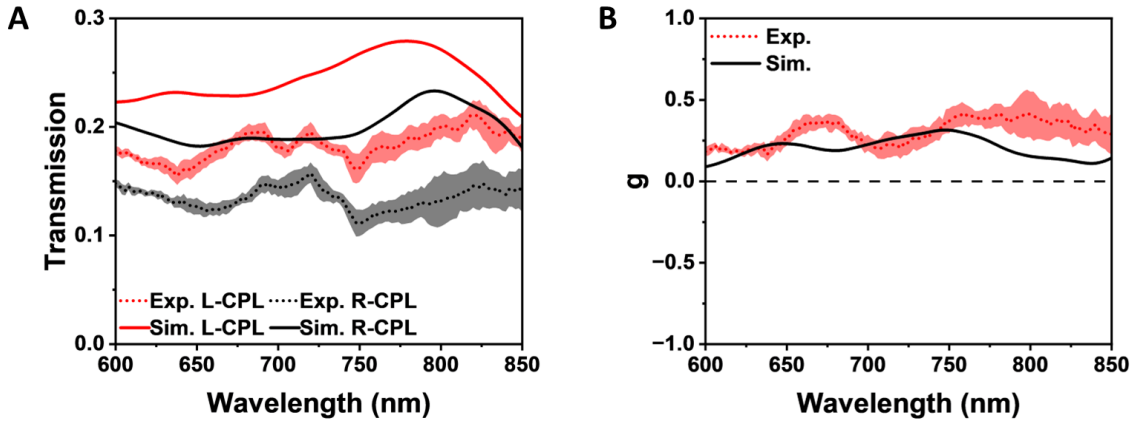

**Figure 5.** (A) Simulated (solid) and experimental (dotted) transmission spectra through a 3D r-AS illuminated by L-CPL (red) and R-CPL (black). (B) Simulated and experimental g-factor of single 3D r-AS. The shaded regions around the experimental curves represent the standard deviation calculated from average spectra. In simulation, numerical aperture (NA) of the collection objective is considered.

**Conclusion**

This work reports suspended 3D AS that achieves single-handed and broadband $\hat{C}$ over an extended spatial volume within its inner cavity in the Vis−NIR range. The successful implementation of the FIB-assisted template stripping technique enables the fabrication of high-definition complex 3D nanostructures. SEM characterization and optical measurement confirm a high agreement between the fabricated 3D r-AS and the intended design. Its broadband chiroptical behavior is experimentally indicated by the consistently sign-preserving g-factor across the operational bandwidth. These results establish a promising foundation for employing the 3D AS as a sieve for trapping enantiomer-coated nanoparticles for ultrasensitive and broadband chiral sensing.

## Methods

### Numerical simulations

The optical response of the 3D AS is simulated using finite-difference time-domain technique (FDTD solutions, 2024 R1.1, Lumerical). The perfectly matched layer boundary condition is involved in all directions. A single AS is excited from the base side by two orthogonally and linearly polarized total-field scattered-field (TFSF) plane wave sources with ±90° phase difference to simulate CPL excitation of different handedness. For linearly polarized light excitation, a single linearly polarized TFSF plane wave source with polarization oriented along the y-axis is used. The mesh resolution is set to 1 nm. A 2D monitor in the y-z plane at x = 0 nm and a line monitor along the z-axis are used to record the near-field optical response of the 3D AS. To record the transmission spectra of the single 3D r-AS, a 2D monitor in the x-y plane placed above the TFSF source box is used. For source normalization, the r-AS is replaced with a hole whose diameter is equal to the r-AS' base diameter, and the corresponding transmission spectra are recorded under identical simulation conditions. The transmission spectrum of the r-AS is normalized by the transmission spectrum of the hole. NA of 0.55 is considered for all far-field calculations.

### Sample fabrication

The 2D array of inverse cones (2 μm base diameter, 1.28 μm depth, 6 μm periodicity) in a Si wafer is defined by $Ga^+$ FIB milling (acceleration voltage 30 kV, beam current 0.23 nA, FEI Helios NanoLab G3UC). 5 nm $SiO_2$ layer is deposited via atomic layer deposition, covering the whole Si surface. This ensures that the subsequently physical-vapor-deposited 80 nm Au film can be easily peeled off during the transfer process. A Si substrate with pre-patterned aperture (40×40 $μm^2$) bought from Silson Ltd is subjected to e-beam evaporated 3 nm Ti followed by thermally evaporated 50 nm Au. This serves as an adhesive holder for the to-be-transferred Au film with cone array. Under an optical microscope, the PMMA-supported Au film with the cone array is peeled off from the template and transferred to the target Si holder, where the Au cone array is aligned to and thus suspended over the pre-patterned square aperture of the Si holder. The PMMA layer is then removed by soaking the Si holder in acetone for several minutes, yielding a clean, hollow and suspended Au cone array. Finally, same FIB machine is used to mill (acceleration voltage 16 kV, beam current 4 pA) the two-arm r-AS pattern into the suspended Au cones, producing well-defined 3D suspended r-AS structures.

### Optical measurements

A supercontinuum white light laser (SuperK COMPACT, NKT Photonics) combined with a tunable filter (SuperK Varia, NKT Photonics) provides broadband illumination in the 600−850 nm range. CPL is generated using a Glan-Laser calcite polarizer (GL10, Thorlabs) in combination with a rotatable achromatic quarter-wave plate (AQWP05M-580, Thorlabs).

Collimated CPL is used to uniformly illuminate the fabricated array from the base side of the r-AS. Given the height of the r-AS, a long working distance (WD) objective lens (50× magnification, NA 0.55, WD 9 mm, LD EC Epiplan-Neofluar, Zeiss) is used to collect the transmitted light. The collected light is then focused by a lens with 300 mm focal length onto a pinhole of 500 μm diameter. This configuration enables spatial filtering in the conjugated image plane to isolate the transmitted signal from one structure. The filtered beam is subsequently collimated by another lens with 300 mm focal length. The collimated beam is then focused onto the entrance slit of a spectrometer (Shamrock 193i, Andor) with a camera (iVac 316, Andor) using a lens with 200 mm focal length. For source normalization, the transmission spectrum of the r-AS is normalized by the transmission spectrum of a hole with the same diameter as the r-AS' base diameter.

**Associated Content**

**Supporting Information**

I. Mathematical derivation

II. $\hat{C}$ of 3D left-handed AS without silica sphere inside

III. $\hat{C}$ of 3D r-AS with silica sphere inside

IV. Optical characterization

V. Single r-AS' simulated transmission spectra under 6×6 array illumination

**Author Information**

**Corresponding Author**

**Min Jiang** − *Leibniz Institute of Photonic Technology, Albert-Einstein-Str. 9, 07745 Jena, Germany; Abbe Center of Photonics, Friedrich Schiller University Jena, Albert-Einstein-Str. 6, 07745 Jena, Germany; ARC Centre of Excellence for Transformative Meta-Optical Systems, Department of Electronic Materials Engineering, Research School of Physics, The Australian National University, Canberra ACT 2600, Australia;* orcid.org/0000-0002-0522-4279

**Email:** min.jiang@leibniz-ipht.de.

**Jer-Shing Huang** − *Leibniz Institute of Photonic Technology, Albert-Einstein-Str. 9, 07745 Jena, Germany; Abbe Center of Photonics, Friedrich Schiller University Jena, Albert-Einstein-Str. 6, 07745 Jena, Germany; Institute of Physical Chemistry, Friedrich Schiller University Jena, Helmholtzweg 4, 07743 Jena, Germany; Research Center for Applied Sciences, Academia Sinica, 128 Sec. 2, Academia Road, Nankang District, Taipei 11529, Taiwan; Department of Electrophysics, National Yang Ming Chiao Tung University, Hsinchu 30010, Taiwan;* orcid.org/0000-0002-7027-3042

**Email:** jer-shing.huang@leibniz-ipht.de.

**Author Contributions**

J.-S.H. conceived the idea and supervised the project. M.J. and A.C. performed the simulations. X.W. developed the complete fabrication protocol and conducted the transfer of Au film with cone array. M.J. and X.W. conducted the FIB milling. A.C. and M.J. performed the optical experiment. M.J. and A.C. analyzed the data. M.J. and A.C. wrote the manuscript. All authors participated in discussions and contributed to manuscript revisions.

**Notes**

The authors declare no competing financial interest.

**Acknowledgments**

The authors gratefully acknowledge the financial support from DFG via IRTG 2675 "Meta-Active" (Project No.: 437527638) and CRC 1375 NOA (Project No.: 398816777). We also thank Birger Steinbach for coating the FIB-milled Si template and the commercially bought Si substrate.

# Supporting Information

# High-definition 3D suspended Archimedean spiral with broadband, spatially extended, and single-handed optical chirality enhancement in Vis-NIR range

Min Jiang,[1, 2, 3, *] Abhik Chakraborty,[1, 2, 4] Xiaofei Wu,[1] Hark Hoe Tan,[3] Jer-Shing Huang[1, 2, 4, 5, 6, *]

[1]Leibniz Institute of Photonic Technology, Albert-Einstein-Str. 9, 07745 Jena, Germany

[2]Abbe Center of Photonics, Friedrich Schiller University Jena, Albert-Einstein-Str. 6, 07745 Jena, Germany

[3]ARC Centre of Excellence for Transformative Meta-Optical Systems, Department of Electronic Materials Engineering, Research School of Physics, The Australian National University, Canberra ACT 2600, Australia

[4]Institute of Physical Chemistry, Friedrich Schiller University Jena, Helmholtzweg 4, 07743 Jena, Germany

[5]Research Center for Applied Sciences, Academia Sinica, 128 Sec. 2, Academia Road, Nankang District, Taipei 11529, Taiwan

[6]Department of Electrophysics, National Yang Ming Chiao Tung University, Hsinchu 30010, Taiwan

Emails: min.jiang@leibniz-ipht.de, jer-shing.huang@leibniz-ipht.de.

**Content:**

I. **Mathematical derivation**

II. **$\hat{C}$ of 3D left-handed AS without silica sphere inside**

III. **$\hat{C}$ of 3D r-AS with silica sphere inside**

IV. **Optical characterization**

V. **Single r-AS' simulated transmission spectra under 6×6 array illumination**

## I. Mathematical derivation

Fundamentally, CD originates from the coupling between induced electric dipole moment (**p**) and magnetic dipole moment (**m**).[1,2] These quantities can be expressed as

$$\mathbf{p} = \tilde{\alpha}\mathbf{E} - i\tilde{G}\mathbf{B}$$
$$\mathbf{m} = \tilde{\chi}\mathbf{B} + i\tilde{G}\mathbf{E} \qquad (1)$$

where, $\tilde{\alpha}$ is the complex electric polarizability, $\tilde{\chi}$ is the complex magnetic susceptibility, $\tilde{G}$ is the isotropic mixed electric-magnetic dipole polarizability, **E** and **B** are the complex electric and magnetic field, respectively. The rate of excitation of a chiral molecule is

$$A^\pm = \frac{\omega}{2}\text{Im}(\mathbf{E}^* \cdot \mathbf{p} + \mathbf{B}^* \cdot \mathbf{m}) = \frac{\omega}{2}\text{Im}[\tilde{\alpha}|\mathbf{E}|^2 + \tilde{\chi}|\mathbf{B}|^2 + i\tilde{G}(\mathbf{E}\mathbf{B}^* - \mathbf{B}\mathbf{E}^*)] \qquad (2)$$

where $A^\pm$ is the absorption rate in left-handed (+) CPL (L-CPL) and right-handed (−) CPL (R-CPL), $\omega$ is the angular frequency. Under the excitation of a time-harmonic electromagnetic field, whose complex electric and magnetic field can be described as $\mathbf{E} = (\mathbf{E}_R + i\mathbf{E}_I)e^{-i\omega t}$ and $\mathbf{B} = (\mathbf{B}_R + i\mathbf{B}_I)e^{-i\omega t}$ respectively. By replacing **E** and **B** by these two relations, eq 2 becomes

$$A^\pm = \frac{\omega}{2}(\alpha''|\mathbf{E}|^2 + \chi''|\mathbf{B}|^2) + \omega G''\text{Im}(\mathbf{E}^* \cdot \mathbf{B}) \qquad (3)$$

The term $\frac{\omega}{2}\chi''|\tilde{B}|^2$ can be ignored because at optical frequencies, materials typically have negligible magnetic response ($\chi'' \approx 0$).[1] Thus the equation can be simplified further as

$$A^\pm = \frac{\omega}{2}\alpha''|\mathbf{E}|^2 + \omega G''\text{Im}(\mathbf{E}^* \cdot \mathbf{B}) \qquad (4)$$

The optical chirality (C) of the illuminating light is[3,4]

$$C = -\frac{\varepsilon_0 \omega}{2}\text{Im}(\mathbf{E}^* \cdot \mathbf{B}) \qquad (5)$$

where $\varepsilon_0$ is the permittivity of vacuum. By using C, eq 4 becomes

$$A^\pm = \frac{2}{\varepsilon_0}(\omega U_e \alpha'' \mp CG'') \qquad (6)$$

where $U_e = \frac{\varepsilon_0}{4}|\mathbf{E}|^2$ is the time-average energy density of the electric field. Note that the physical meaning of the first term is the conventional dipolar absorption and the true origin of CD is the second term. In other words, to enhance the absorption difference, one needs to engineer the field to maximize C.

## II. $\hat{C}$ of 3D left-handed AS without silica sphere inside

In the main text, we have discussed $\hat{C}$ calculated along the central axis of the 3D r-AS spanning a continuous wavelength range of 600−850 nm. To have a complete view of 3D AS, the oppositely handed structure's behavior is investigated here. We conduct numerical simulations of the 3D left-handed AS (l-AS) also in the absence of any silica sphere. The numerical simulation is performed using the finite-difference time-domain technique (FDTD Solutions, 2024 R1.1, Lumerical). The perfectly matched layer (PML) boundary condition is applied in all directions. We use two orthogonally and linearly polarized total-field scattered-field (TFSF) plane wave sources with ± 90° phase difference, illuminating the l-AS from its base side. The mesh step is uniformly set to 1 nm in all directions within the volume enclosing the l-AS. In the case of linearly polarized light excitation, a single linearly polarized TFSF plane wave source with polarization oriented along the y-axis is used with every other parameter kept the same as that in the CPL condition.

A line monitor (black line in **Figure S1A**) aligned along the central axis (z-axis) records the electric and magnetic field components over the entire operational wavelength range. Under L-CPL illumination (**Figure S1A**), a strong and single-handed $\hat{C}$ is observed across wavelengths ranging from 600 nm to 850 nm and within the spatial range of z = 800−1200 nm. $\hat{C}$ peaks near the tip (z = 1100−1200 nm) with the value up to 25. Notably, this result matches the distribution shown in **Figure 3A**, with the only difference being the sign inversion. In contrast, under R-CPL illumination (**Figure 3B**), which is helically opposite to the handedness of the structure, both handedness of $\hat{C}$ appear, especially near the tip, and the overall $\hat{C}$ is significantly reduced. Additionally, as shown in **Figure S1C**, single-handed and broadband $\hat{C}$ is also observed but more confined near the tip under linearly polarized light illumination.

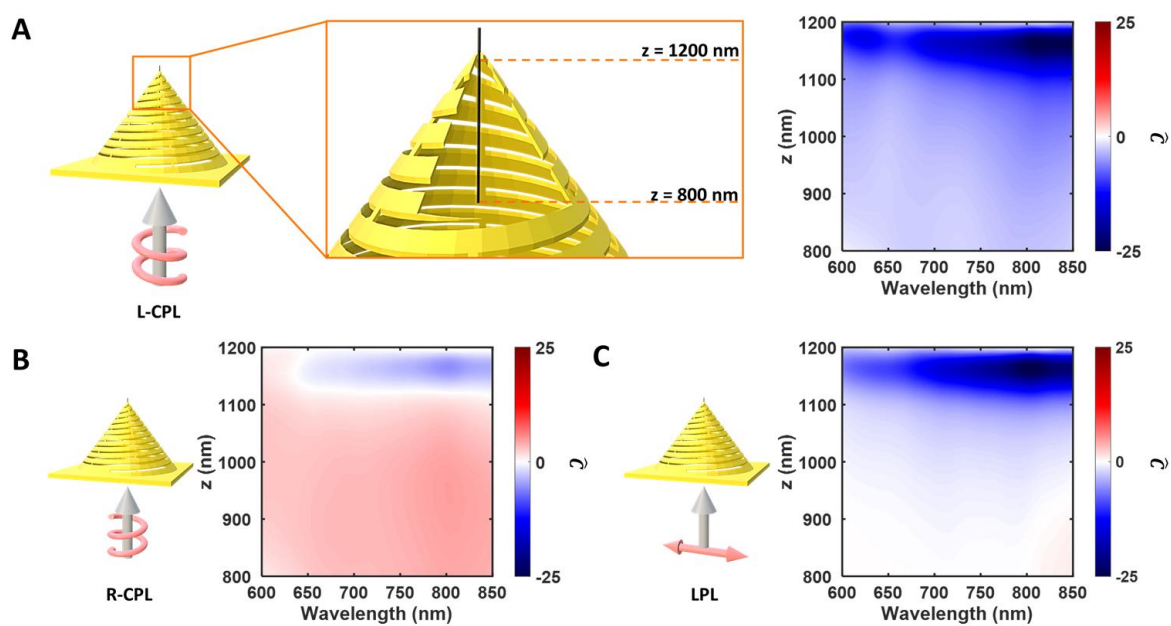

**Figure S1.** $\hat{C}$ of 3D suspended Au l-AS without silica sphere inside calculated along z-axis line monitor under (A) L-CPL, (B) R-CPL, and (C) linearly polarized light excitation.

The full mapping of $\hat{C}$ in both 3D r-AS and 3D l-AS suggests that two broadband and single-handed $\hat{C}$ with same amplitude but opposite sign spanning across the functional region can be generated when oppositely handed AS structures helically match oppositely handed CPL excitation. As $C$ is proportional to CD, this is extremely useful for CD measurement considering that absorption difference is typically smaller than the overall absorption of the sample by some orders of magnitude. Therefore, our 3D AS shows great potential as a platform for chiral sensing.

### III. $\hat{C}$ of 3D r-AS with silica sphere inside

To validate the idea of using 3D r-AS to trap the functionalized nanoparticle within the functional region for chiral sensing, we perform simulations under the same conditions as described above, now incorporating a 100 nm-diameter silica sphere positioned near the tip.

Due to the geometric constraints imposed by the size of the silica sphere and the height-dependent inner diameter of the r-AS, the top surface of the sphere cannot reach z = 1200 nm. The positions of top and bottom surface of the silica sphere are indicated by turquoise dashed lines. We then calculate Ĉ through the line monitor aligned with z-axis. Under R-CPL and linearly polarized light illumination, we observe single-handed, broadband and spatially extended Ĉ, which does not occur under L-CPL illumination. Notably, the presence of the nanoparticle induces no noticeable change in the Ĉ distribution compared to the case without any nanoparticle (**Figure 3**). This solidifies the potential of the 3D r-AS to function as a sieve for chiral sensing at low analyte concentration.

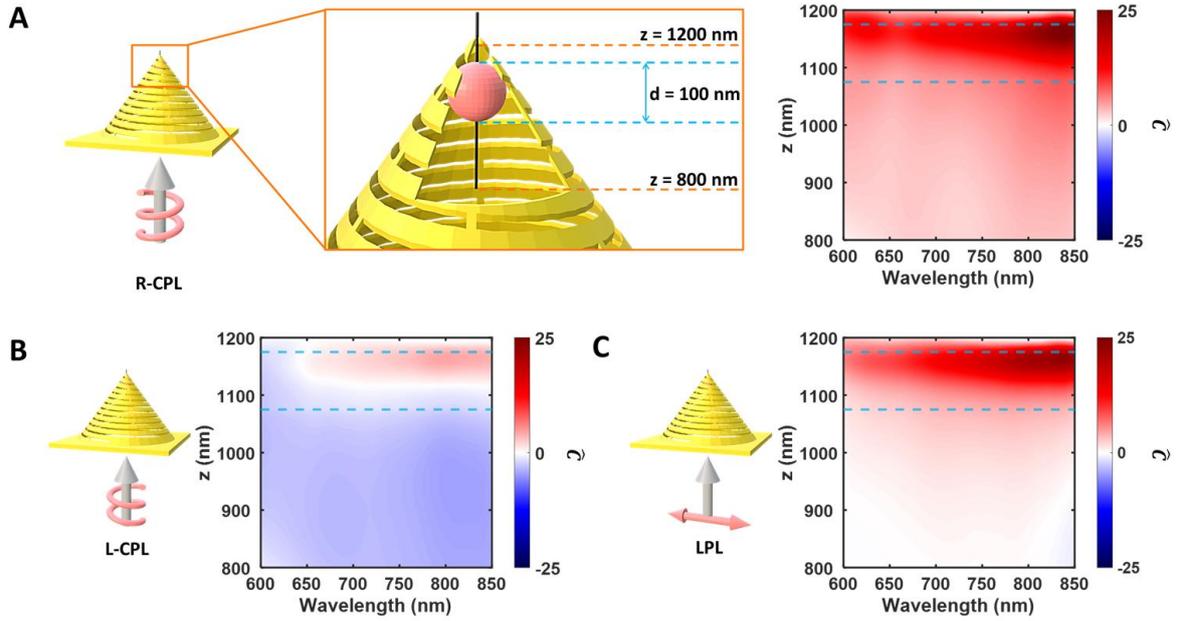

**Figure S2**. Ĉ of 3D suspended Au r-AS with silica sphere inside calculated along z-axis line monitor under (A) R-CPL, (B) L-CPL, and (C) linearly polarized light excitation.

## IV. Optical characterization

The fabricated sample consists of 6×5 3D r-AS array and an adjacent 6×1 hole array with hole diameter matching the r-AS base diameter. The entire 6×6 array rests on a suspended Au film. Here, the hole serves as a reference source for normalization. **Figure S3** shows the scanning electron microscope (SEM) image of this sample used for optical characterization. We select the second column (marked with red box) from the left side of the array and measure the signal from each individual r-AS in the column. We normalize the results against the source spectrum obtained from the hole (marked with blue box) in the rightmost column. The transmitted intensity collected from each r-AS under a certain handedness of CPL is normalized by the transmitted intensity collected from a hole excited by the same CPL handedness. The transmission (T) spectrum of a 3D r-AS is calculated using the following equation:

$$T = \frac{I_{r-AS} - I_{background}}{I_{source} - I_{background}} \quad (1)$$

Where, $I_{r-AS}$ and $I_{source}$ are the transmitted intensity from a single 3D r-AS and a single hole respectively, and $I_{background}$ is the contribution from the environment.

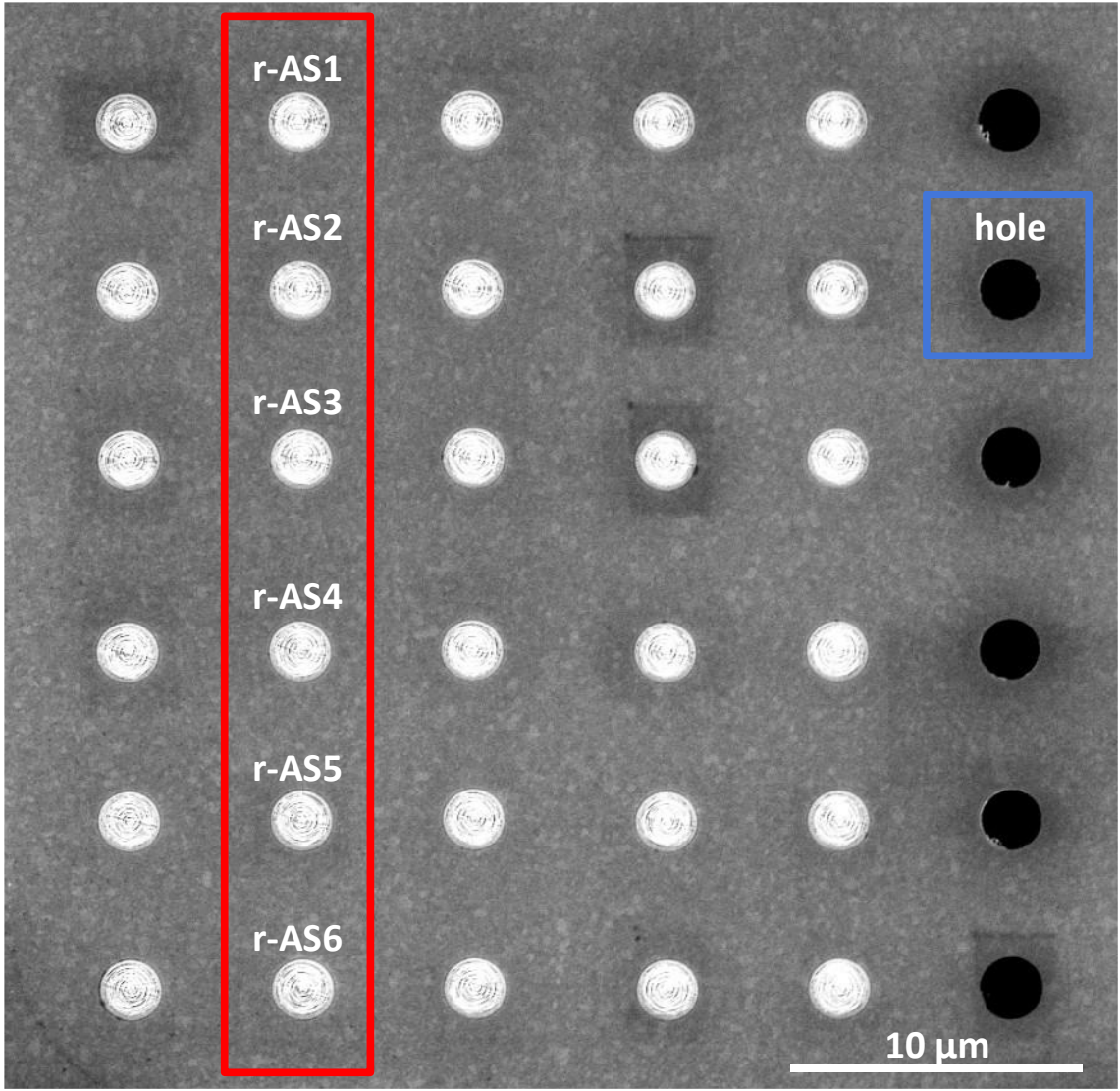

**Figure S3.** SEM image of the fabricated 6×6 array (6×5 3D r-AS array and 6×1 clear hole array) used for transmission measurement. Transmission spectra of r-AS1−r-AS6 (marked with red box) in the second column (from left to right) are measured. The transmission spectrum of each r-AS under the illumination of a particular helicity is normalized as $\frac{I_{r-AS}-I_{background}}{I_{source}-I_{background}}$, where $I_{r-AS}$ and $I_{source}$ are the transmitted intensity spectra from one r-AS and one hole (marked with blue box) respectively. $I_{background}$ is the contribution from the environment.

We use a home-built transmission microscope to characterize the chiroptical response of the sample shown in **Figure S3**. This system is constructed based on the 4f optical imaging system which helps to isolate individual structure's signal through a pinhole positioned at the plane where the conjugated image of the sample plane is created. The optical setup is sketched in **Figure S4**. A supercontinuum white light laser (SuperK COMPACT, NKT Photonics) combined with variable bandpass filter (SuperK Varia, NKT Photonics) provides broadband illumination covering the 600−850 nm spectral range. CPL is generated using a Glan-Laser calcite polarizers (GL10, Thorlabs) in combination with a rotatable achromatic quarter-wave plate (AQWP05M-580, Thorlabs). Collimated light illuminates the entire array uniformly. This is done as opposed to employing focused incidence to reduce the intensity of light at every point in the sample plane and avoid the risk of photodamage to the delicate 3D structures. The light transmitted through the sample is collected using an objective lens (LD EC Epiplan-Neofluar, Zeiss) with a long working distance (9 mm), 50× magnification and 0.55 numerical aperture (NA) and then focused by a lens with a focal length of 300 mm onto a 500 μm diameter pinhole positioned at the conjugated image plane. This configuration enables spatial filtering to isolate the transmission signal from a single 3D r-AS or hole. The transmitted beam is subsequently collimated using another lens with a focal length of 300 mm, followed by a lens with a focal length of 200 mm that focuses the light onto the entrance slit of a spectrometer (Shamrock 193i, Andor) with camera (iVac 316, Andor).

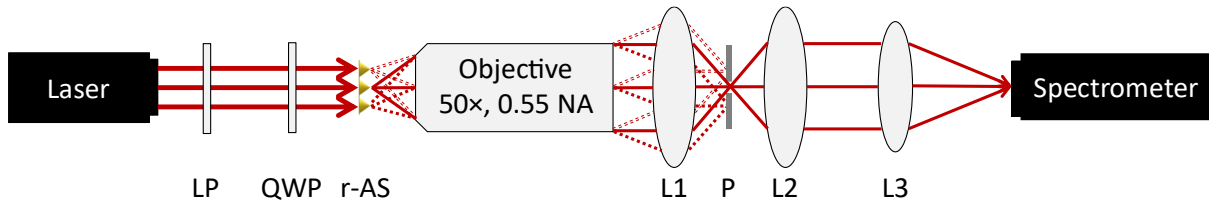

**Figure S4.** (A) Schematic of the optical setup to measure the transmission spectra through the fabricated 3D r-AS. LP: linear polarizer; QWP: quarter-wave plate; L1: lens 1 with a focal length of 300 mm; P: 500 μm diameter pinhole; L2: lens 2 with a focal length of 300 mm; L3: lens 3 with a focal length of 200 mm.

**V. Single r-AS' simulated transmission spectra under 6×6 array illumination**

Influence of simultaneously illuminating the whole 6×6 array (6×5 3D r-AS array and the adjacent 6×1 hole array) is studied using finite-difference time-domain (FDTD solutions, 2024 R1.1, Lumerical). The PML boundary condition is involved in all directions to cover the entire simulation region. Two separate simulations are conducted to account for L-CPL and R-CPL excitation. In both cases, the entire array is excited from the r-AS' base side by two linearly and orthogonally polarized TFSF plane wave sources with phase difference of ±90°. The mesh resolution is set to 5 nm in all directions covering each structure. Despite the illumination of the entire array, care is taken to make sure that the transmission spectrum recorded by and computed at the 2D monitor positioned in the x-y plane above the TFSF source box only collects the response of one structure at a time. The transmission spectrum of each structure marked in **Figure S3** is separately computed for both L-CPL and R-CPL excitation considering NA of 0.55. The transmission spectrum for each r-AS computed under a certain handedness of CPL excitation has been normalized by the transmission spectrum for the hole computed under the same handedness of CPL excitation in order to replicate the experimental condition. **Figure S5** shows the average normalized transmission spectra simulated under L-CPL and R-CPL excitation. Clearly, with the illumination of whole 6×6 array, some spectral features are revealed in the single r-AS transmission that are similar to those depicted in the experimental transmission spectra provided in **Figure 5A**. Therefore, despite the fact that spatial filtering is done via a pinhole to isolate the signal from a single structure in our experiment, some inter-structural coupling and interference may still occur if the entire array is illuminated at once.

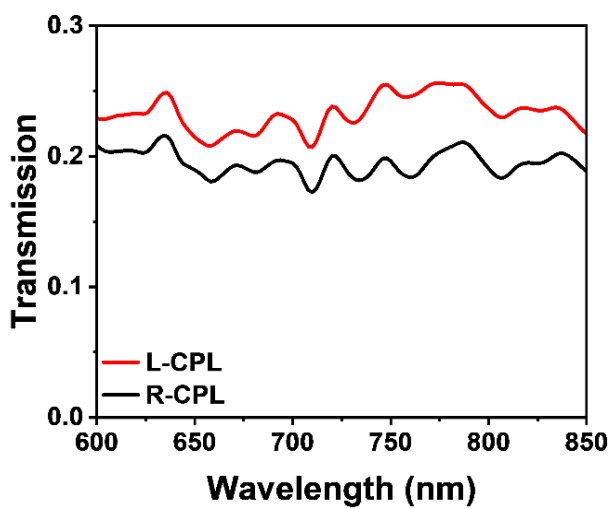

**Figure S5.** Simulated average normalized transmission spectra corresponding to the response of the individual 3D r-AS under L-CPL and R-CPL illumination of 6×6 array (6×5 3D r-AS array and 6×1 hole array). Here, the hole serves as a reference source for normalization. Each structure marked in **Figure S3** is separately computed under both L-CPL and R-CPL illumination, considering a numerical aperture of 0.55. The normalized transmission spectrum for each r-AS is then calculated as $I_{r-AS}/I_{source}$, where the numerator and denominator correspond to transmissions under the same handedness of CPL.